# IEEE Copyright notice:

This paper is accepted and will be published in the Proceedings of the 2019 IEEE PES General Meeting, August 4-8 2019, Atlanta, GA, USA.

# STATCOM Performance Evaluation Using Operation Data from Digital Fault Recorder

Duotong Yang, Hung-Ming Chou, Kyle Thomas
Dominion Energy
Richmond, VA, USA

Sergey Kynev
SIEMENS
Raleigh, NC, USA

Rebecca Rye
Virginia Tech
Blacksburg, VA, USA

*Abstract*—**Static Synchronous Compensators (STATCOMs) are being employed by Dominion Energy to control voltage and enhance system stability. Due to the complexity of the control systems, operational modes, and nonlinearities, it is essential to evaluate STATCOMs' behavior to ensure their correct and proper response to dynamic events such as line faults, generator trip, or load rejection throughout the grid. This procedure brings benefit to device management by identifying potential equipment problems and improve dynamic model for simulation study. The proposed framework utilizes operation data collected in Digital Fault Recorder (DFR) to evaluate STATCOM's response to dynamic system events. One of the challenges on getting accurate model response comes from STATCOMs with automatic gain adjustment feature. This feature actively measures the external system's Thevanin impedance and accordingly changes the STATCOM gain. Therefore, when trying to recreate the field measurements in the simulation environment, it is necessary to match the short circuit level (SCL) of the external system equivalent. This paper presents how these issues have been solved in order to evaluate STATCOM's performance. Finally, the effectiveness of proposed performance evaluation method is validated based on two actual events caused by faults.**

## INTRODUCTION

In recent decade, the progress achieved in semiconductor technology prompted the use of high speed self-commutated switches such as IGBTs and GTOs for high voltage power delivery [1]. Flexible AC Transmission System (FACTS) devices based on IGBTs and GTOs have been widely deployed across power network to increase controllability and transfer capacity [2] [3]. Static Synchronous Compensator (STATCOM) came in service for the first time since 1991 in Japan [4]. Since then, STATCOM have been deployed in commercial operation to provide voltage control. Comparing with Capacitor bank, STATCOM provides more flexibility and faster response on reactive power support. In addition, STATCOM offers a unique advantage of generating both inductive and capacitive power with only one single Modular Multilevel Converter (MMC) so that it requires less space when comparing with thyristor based Static VAR Compensator (SVC) [5].

To guarantee the reliability and accuracy of STATCOMs' performance, evaluating STATCOM's performance against vendor's validated model after commissioning or controller update is mandatory [6] [7]. However, the complexity of STATCOM control system and different control modes present challenges for operator and field engineers to verify devices' behavior. Besides, most of the control algorithms are protected by vendor's IP rights. Hence, a solution that can evaluate FACTS devices performance to ensure their correct control response is demanding. In recent decades, researchers and engineers have proposed several methods to evaluate STATCOM or other FACTS devices' performance using Electromagnetic transient analysis tools (EMT-type) model in real time or offline. For example, the most commonly used method is to compare field measurement against EMT model's response during commissioning's start-up sequence to ensure devices' reliability and accuracy [8].

Besides evaluating start-up sequence, engineers are also curious about devices' performance under disturbances, especially scenarios during line faults, generator tripping, and load rejection. However, very few researches are conducted to provide FACTS performance evaluation or model validation with operation data. One of the challenges is the difficulty of replicating external system condition. For example, STATCOM model with automatic gain adjustment cannot change the gain setting without simulating its external system condition.

The methodology introduced in this paper provides a cost-effective solution to evaluate STATCOM's performance. The method utilizes vendor's three phases EMT-type model in PSCAD to provide STATCOM's expected response. Field operation data recorded in DFR is also used to provide model's source voltage and field reactive power measurements. To automate the evaluation procedure, an automation framework is developed to extract EMS settings for STATCOM and DFR measurements. The proposed framework is also able to simulate AC external system's short circuit level which represents system's external network condition so that STATCOMs with Auto-Gain change can adjust their gain accordingly.

## CONFIGURATION OF STATCOM

Fig.1 shows a schematic diagram of Dominion's STATCOM system connected with transformer and AC grid. This simplified STATCOM model shows a VSC for AC voltage/reactive power control and DC Capacitor that have a relatively constant DC voltage with minimum ripple. Modern STATCOMs based on MMC technology contains of many individual submodules with a smaller DC Capacitors within each submodule.

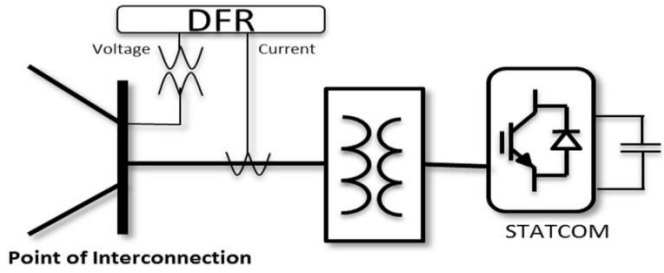

Figure 1 STATCOM configuration

### A. STATCOM Control Schemes

Modern STATCOMs are able to operate under two major control modes: voltage control mode (VCM) and Q control mode (QCM). VCM provides capacitive or inductive reactive power to maintain voltage by referring to voltage set point and reactive power output. The control can be depicted as the following equation.

$$V_{act} = V_{ref} - slope \cdot \frac{Q_{act}}{Q_{nominal}} \quad (1)$$

Where $V_{act}$ and $V_{ref}$ represents actual and reference voltage respectively. $slope$ indicates STATCOM's droop control slope. $Q_{act}$ represents actual reactive power output, while $Q_{nomial}$ denotes its nominal reactive power value. According to the blue curve in Figure 2, STATCOM is providing capacitive power if operating point is at the left hand side. When operating point is at the right hand side, STATCOM will provide inductive power to bring back voltage.

If STATCOM is operating under VCM and QCM at the same time, STATCOM will quickly operate along the blue curve when disturbance occurs. As soon as it reaches the intersected point with system load line (green), STATCOM will start bringing back the operating condition to the Q set point $Q_{ref}$. It is noted that QCM takes longer time to operate than VCM, therefore, VCM is the key control function during transient disturbance.

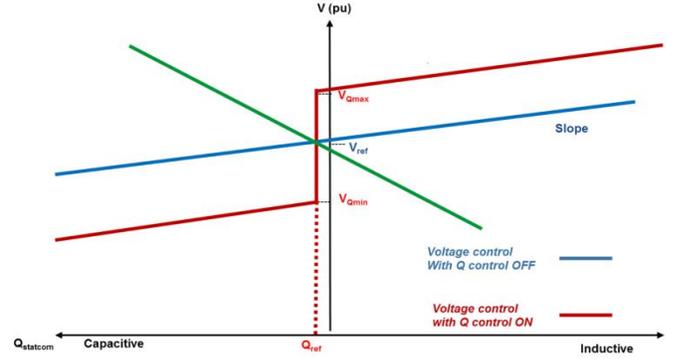

Figure 2 STATCOM's QV Control Mode

Voltage control schemes of STATCOM can be varied from different vendors and utilities' specific requirements. For area with high three phases unbalance, utilities will prefer vendors to provide STATCOM with negative sequence voltage control. For region with risk of large short circuit level (SCL) changes, gain adjustment is needed. The higher SCL indicates the stronger AC system. STATCOM's gain setting can be maintained in a relatively higher value when system is strong. A lower SCL indicates a weaker AC system and thus the gain will be adjusted to a lower value to prevent overshoot and system instability condition. Gain adjustment methods can be divided into passive and automatic.

Passive gain adjustment reduces gain when controller detects multiple consecutive changes in control current signal. The gain will be reduced stepwise in a certain percentage until stability is reached. When disturbance is cleared, the gain will be recovered to its default setting. One of the drawbacks is that passive method provides limit number of gain values for adjustment. Besides, the gain setting cannot increase when SCL changes from low to high. In order to solve this problem, automatic gain adjustment is introduced. Figure 3 shows a field measurement of reactive power and gain change after automatic gain adjustment. As can be seen, STATCOM injects reactive power to AC grid and measures its SCL by calculating dQ/dV. Based on the computed dQ/dV, STATCOM can adjust its gain accordingly. Automatic gain adjustment can be activated cyclically several times a day.

In this paper, the STATCOM under study deploys automatic gain adjustment function. This function is also implemented in its PSCAD model and therefore its gain cannot be adjusted manually. In order to set the gain to its pre-fault value, a scheme is presented in this paper to provide pre-fault system SCL to PSCAD model so that its gain can be adjusted automatically.

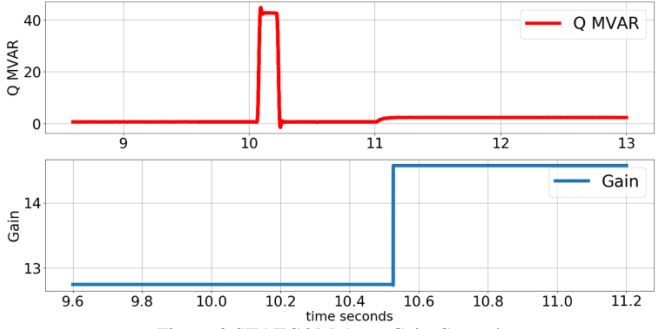

Figure 3 STATCOM Auto-Gain Control

### B. Measurements of STATCOM

STATCOM's setting of voltage reference, reactive power reference, slope, gain, etc., are stored in EMS system.

When disturbance occurs, STATCOM's transient measurements are captured in DFRs with sampling rate of 9600 per second. As shown in Figure 1, the voltage and currents used for voltage and Q control are measured at the point of interconnection. The reactive power output is computed based on the primary side of voltage and current. Therefore, the primary side measurements are the major concern in performance evaluation. Besides AC voltage and current signals, DC voltage measurements and control currents are also stored in DFRs.

PERFORMANCE EVALUATION SETUP

Figure 4 shows the detail procedure of proposed STATCOM performance evaluation scheme.

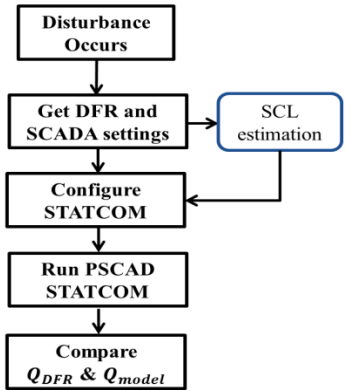

Figure 4 Workflow of Performance Evaluation

As illustrated in Figure 4, STATCOM's DFR is activated as soon as disturbance is captured. Then automatic system starts extracting DFR measurements and STATCOM's EMS pre-event settings when DFR data is ready to process. The pre-fault SCL is estimated based on the given EMS gain value. Then based on the estimated SCL and pre-fault EMS settings, STATCOM's PSCAD model is configured. After model's configuration, PSCAD starts running by injecting DFR primary side voltage at point of interconnection while providing corresponding reactive power response. Finally, simulated reactive power output and DFR reactive power measurements are compared to provide performance evaluation report.

### C. Gain Setting

To set the gain value for STATCOM with automatic gain adjustment, the corresponding external SCL needs to be estimated. The relationship between SCL and gain value can be found by varying the external impedance and the slope of droop control in STATCOM's PSCAD model under nominal voltage. According to the EMS setting, the slope of STATCOM's droop control is fixed. In transmission system, the resistance of external impedance is negligible in comparison to the reactance. Hence, the resistance of external impedance can be fixed with a relatively small number in PSCAD model. Therefore, the gain value can be determined based on the external reactance only.

Several gain values and their corresponding external reactance as well as SCLs are shown in Table 1. Based on the plotting between actual gain and reactance value (Figure 5), it can be observed that their relationship shows a nonlinear behavior and can be represented by a polynomial regression model. Figure 5 shows the difference between the curve representing the relationship and the polynomial fitted curve. As can be seen, the polynomial curve accurately portraits the actual relationship between gain and external reactance before gain value reaching 25. It has to mention that the polynomial curve and

**Table 1** are project specific and only apply to the STATCOM under studied in this paper only.

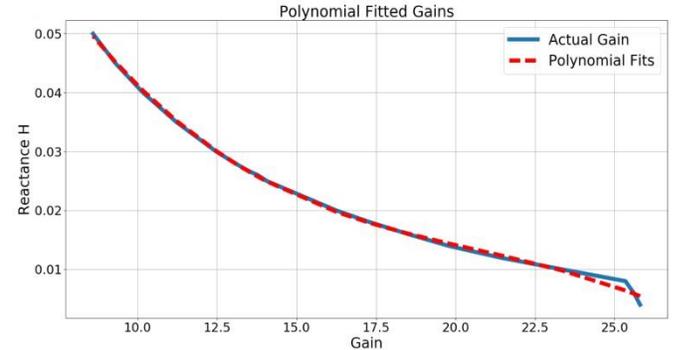

Figure 5 Actual Gain and Polynomial Fits

### D. DFR Data Injection

As introduced in 0, voltage and current are collected at the point of interconnection in DFR. Therefore, the DFR voltage data can be injected at the point of interconnection in PSCAD model. As shown in Figure 6, the injected voltage source is playing as an "infinite-bus" machine in PSCAD. STATCOM's simulated voltage and frequency are forced to follow very closely to the recorded voltage and frequency.

Table 1 Gain values versus external reactance

| L (H) | ΔQ/ΔV | Vendor's Lookup Table Gain | PSCAD gain |
|---|---|---|---|
| 0.01 | 14.3 | 22.44<x<23.10 | 23.35 |
| 0.02 | 7 | 15.69<x<16.28 | 16.25 |
| 0.025 | 5.342857143 | 12.64 <x<14.26 | 14.06 |

If vendor's model is accurate, the simulated reactive power response should be very close to the recorded reactive power. It has to mention that the automatic gain adjustment function is activated after PSCAD model's start sequence. Therefore, a 10 seconds steady state voltage with nominal value is played before the DFR recorded transient voltage so that the gain adjustment can be completed. It is noted that the external reactance is used for automatic gain adjustment only so that when the gain is changed the estimated external reactance needs to be bypassed. This is because DFR is measuring the voltage and current at the point of interconnection. Thus, the collected DFR voltage measurement has included the external system information and the estimated external reactance will not be needed.

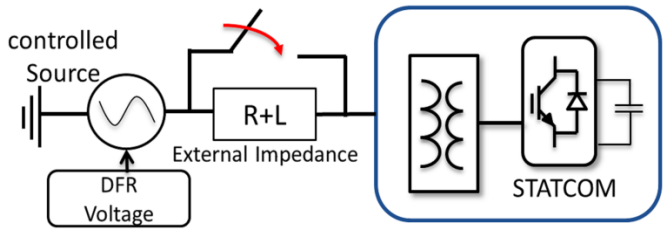

Figure 6 Running PSCAD Model

STATCOM PERFORMANCE EVALUATION

A. *System Setup*

To provide reliable electricity delivery for the customers in Virginia Beach area, Dominion Energy deploys four STATCOMs with same capacity and settings to provide reactive power support. The STATCOM under study in this paper is one of them. The STATCOM is providing reactive power +/- 125 MVAR at 230 kV bus. The STATCOM is running under VCM and QCM at the same time. To have better reactive power margin, $Q_{ref}$ is set as 0 MVAR. The droop control slope is set as 1%.

The STATCOM model is built based upon PSCAD. The automation system running performance evaluation is written in python. The server hosting the automation system is running with 16 GB RAM and an Intel(R) i7-6820HQ CPU @ 2.7 GHz.

B. *Performance Evaluation for Case March 12 2018*

Figure 7 shows a disturbance on March 12 2018. A single line to ground fault occurred and triggered the DFR. As can be seen, there is a significant dip at the voltage measurement. The STATCOM quickly react to the event by providing capacitive reactive power. As soon as the fault is cleared, the STATCOM reduced the reactive power supply to mitigate overvoltage. The whole event data is less than 2 seconds.

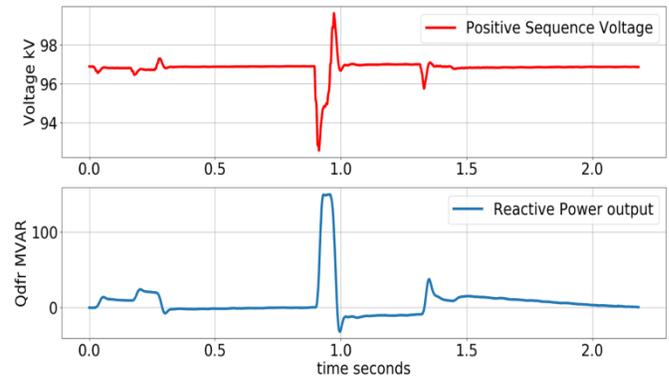

Figure 7 Voltage and reactive power measurements in DFR March 12 2018

The EMS system indicates that the pre-fault gain setting is 14.2. Based on the polynomial model, the external impedance can be found by the given gain value. With gain equals to 14.2, the external reactance L is set as 0.02454 H. The gain value of the PSCAD model changed from 12.75 to 14.57 when automatic gain adjustment is completed.

The performance evaluation result is shown in Figure 8. As can be seen, the trend of the reactive power output stored in DFR is the same as the simulated reactive power response. Besides, Table **2** illustrates that the maximum PSCAD reactive power response is close to the actual field measurement in PSCAD. Therefore, it can be concluded that the STATCOM is providing the same performance as what it is expected to be.

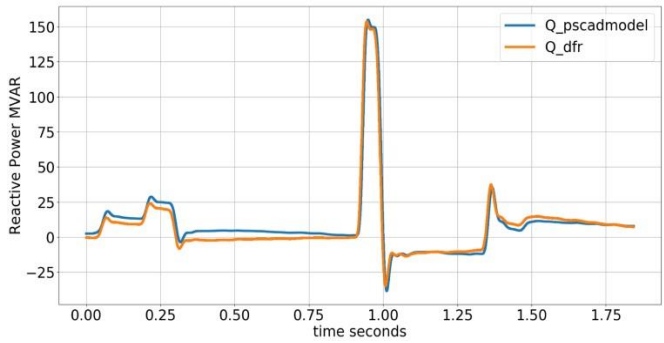
Figure 8 Performance evaluation result March 12 2018

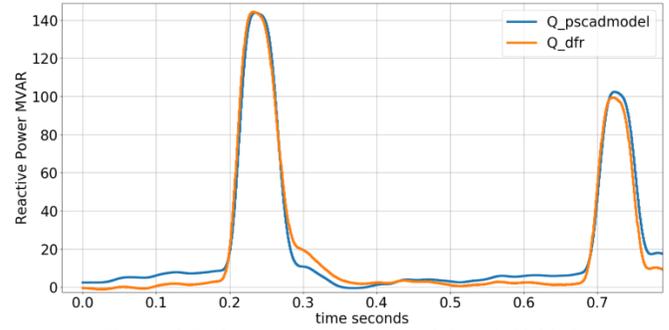
Figure 10 Performance evaluation result March 20 2018

*C. Performance Evaluation for Case March 20 2018*

Another single line to ground fault event happened on March 20 2018 is illustrated in Figure 10. Two significant voltage drops are observed. In this event, the recloser opened the circuit when the fault occurred in the beginning. Then it operated to reclose the circuit, but the fault was still present and hence the voltage dropped again. Finally, the recloser opened the circuit at the second time to clear the fault.

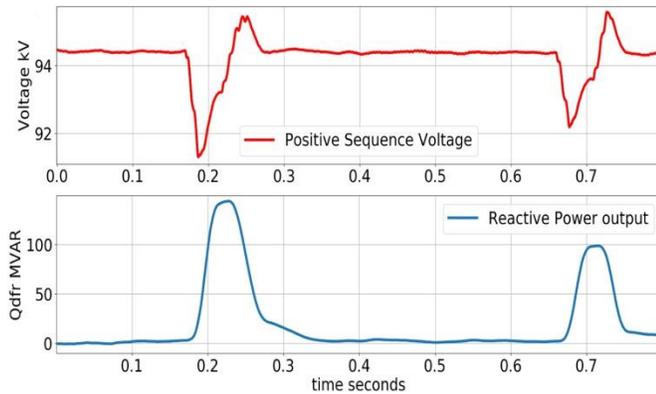
Figure 9 Voltage and reactive power measurements in DFR March 20 2018

In this case, the EMS system shows that the STATCOM gain is set as 12.8. The external reactance is yielded as 0.0289 H based on the polynomial model. The PSCAD gain value is slightly reduced from 12.75 to 12.70 with automatic gain adjustment system.

The result of performance evaluation is shown in Figure 10. Similar as the first case, the first reactive power output shows a good match between simulated response and field measurement.

Table 2 Difference between $Q_{dfr}$ and $Q_{pscad}$

| Case | Maximum $Q_{dfr}$ MVAR | Maximum $Q_{pscad}$ MVAR |
|---|---|---|
| 20180312 | 153.53 | 154.77 |
| 20180320 | 144.29 | 143.824 |

CONCLUSION

In this paper, a systematical approach is proposed for STATCOM performance evaluation against vendor's validated model after commissioning and/or controller update. The proposed methodology is realized through an automation framework. This framework is able to estimate SCL for STATCOM and configure the PSCAD model based on EMS settings.

Testing results show that the proposed solution is able to provide reactive power response which closely matches to the reactive power measurements from the DFR when playing back operation data. This study demonstrates that utilities can use this framework to conduct FACTS devices performance evaluation.